\documentclass{elsart}
\usepackage{amssymb}
\def\be{\begin{equation}}
\def\ee{\end{equation}}
\def\ba{\begin{eqnarray}}
\def\ea{\end{eqnarray}}

\def\th{$^{th}$}
\begin{document}

\begin{frontmatter}
\title{Photographing the wave function of the Universe}
\author{Carlo R. Contaldi, Rachel Bean, and Jo\~ao Magueijo}
\address{
Theoretical Physics, The Blackett Laboratory, \\
Imperial College, Prince Consort Rd., London, SW7 2BZ, U.K.}
\maketitle
\date{}

\begin{abstract}
We show that density fluctuations in standard inflationary scenarios may
take the most general non-Gaussian distribution if the wave function
of the Universe is not in the ground state.
We adopt the Schr\"odinger picture to find a remarkable similarity
between the most general inflaton wavefunction 
and the Edgeworth expansion used in probability theory. 
Hence we arrive at an explicit relation between the cumulants
of the density fluctuations and the amplitudes 
or occupation numbers of the various energy eigenstates. 
For incoherent superpositions only even cumulants may be non-zero,
but coherent superpositions may generate non-zero odd cumulants
as well. Within this framework measurements of cumulants in 
Galaxy surveys directly map the wavefunction of the Universe.
\end{abstract}
\end{frontmatter}

\section{Introduction}
Gaussian distributed density fluctuations are seen as a 
hallmark of the inflationary paradigm~\cite{guth81,linde,as,linde1}. 
In addition,
Gaussianity greatly simplifies calculations as only one set of 
parameters, the power spectrum, needs to be calculated. 
For these reasons, Gaussianity is
commonly assumed as a matter of course. However in non-minimal
models of inflation \cite{salopek,mukh,peebles} it is possible to generate
non-Gaussian statistics. The magnitude \cite{pierpaoli99}
and type of departures from Gaussianity in these models 
appear to be restricted; in the simplest cases fluctuations are
the square, or a power, of a Gaussian random field \cite{peebles}.
Furthermore, there have been a number of recent claims of detection of
a non-Gaussian signal in the COBE data \cite{us,nov,nvs,pando,teg,band,int}.
It is therefore of interest to explore further non-Gaussian inflationary
models. For instance, one may wonder whether inflation can ever 
produce fluctuations with the most general probability distribution.
It could be the case that some types of non Gaussianity are barred
from even the most contrived inflationary models.

In this paper we consider standard inflationary models, and revisit the
derivation of the Gaussianity of density perturbations in these
models. Microphysical fluctuations in the inflaton field then
satisfy very approximately a linear harmonic oscillator equation.
These should be subject to quantisation, with the common assumption 
being that the field lies in the ground state, which for an harmonic
oscillator has
a Gaussian wave function (eg. \cite{qm}). These microphysical 
fluctuations are then
stretched by inflationary expansion, reaching what we nowadays
call cosmological length scales. The square of their wavefunction
provides their probability distribution, and since the square of a
Gaussian is a Gaussian, these fluctuations have Gaussian 
statistics\footnote{Contrary to a popular myth, the Gaussianity 
of standard inflation's fluctuations
is not due to the central limit theorem.}.

Let us now assume that inflation is triggered by quantum cosmology.
However let us
assume nothing from quantum cosmology (cf. \cite{hartle,jonathan})
about the initial form of the inflaton's wavefunction, $\psi$,
allowing it to be given by
the most general superposition of energy eigenstates 
(a situation already considered in \cite{nv,lesg}). The
wavefunctions $\psi_n$ of these eigenstates are generally a Gaussian 
multiplied by a Hermite 
polynomial $H_n$ \cite{qm}. The square of the wave function, $|\psi|^2$,
is the probability distribution of fluctuations emerging from such a scenario. 
The distribution we have just found has a striking resemblance to
the Edgeworth expansion used in probability theory \cite{Kendall}.
The Edgeworth expansion parameterises the most general
distribution, and takes the form of a Gaussian multiplied by a
series of Hermite polynomials. The coefficients in this series are
polynomials in the cumulants of the distribution. Hence we may
write a simple expression relating the components of $\psi$
and the cumulants of the density fluctuations' distribution.
The purpose of this Letter is to establish this interesting connection
between observational and quantum cosmology. 

The plan of this Letter is as follows. In section~\ref{wave} we
review inflationary fluctuations, purposefully 
adopting the Schr\"odinger picture. In section~\ref{edge} we review
the other tool used in this Letter -- the Edgeworth
expansion. Then in sections~\ref{res1} and~\ref{res2}
we establish the connection between the amplitudes of the various modes in the
wavefunction and the cumulants of the distribution of the ensuing
density fluctuations. We find that
for an incoherent superposition of eigenmodes one may only obtain
non-Gaussian distributions with non-zero even cumulants. However
for coherent superpositions it is possible to generate the most
general non-Gaussian distribution. 
Our results therefore show that if we take standard inflation
for granted, we can in principle experimentally map the wavefunction of 
the Universe.  The implications of such a measurement are briefly
discussed in a closing section.

\section{Inflationary fluctuations in the Schr\"odinger picture}
\label{wave}
We first review the formalism used for describing inflationary fluctuations
in the Schr\"odinger picture (eg. \cite{nv,pi}). 
We assume that inflation arises from a single 
scalar field, the inflaton field $\phi$. 
We write $\phi$ as an homogeneous background $\bar{\phi}$ plus a
perturbation $\delta\phi$:  
\be
\phi(\mathbf{x},t)=\bar{\phi}(t)+\delta\phi(\mathbf{x},t) \  \
\ee
with $\delta\phi\ll\bar{\phi}$. It is standard to decompose
the perturbation $\delta\phi(\mathbf{x},t)$ into Fourier modes
$\phi_{\bf k}$. In terms of these, the energy density fluctuations
are given by \cite{nv,pi}:
\be
\delta\rho(\mathbf{k},t) = \dot{\bar\phi}(t)\dot\phi_{\bf k}(t)
\ee
Density fluctuations are therefore 
proportional to $\dot\phi_{\bf k}$, which  in turn is
proportional to $ \phi_{\bf k}$. Therefore deviations
from Gaussianity in $\delta \rho$  may be studied directly
in terms of the statistics of $ \phi_{\bf k}$ (though this
is not true in the models considered in \cite{barrow}). 

At early times we can adopt the free field theory limit and the modes 
$\phi_{\bf
k}$ are then completely decoupled and behave as individual harmonic
oscillators with a time-dependent mass $m(t)$ \cite{nv,pi}. 
The form of $m(t)$ is dictated by the form of the potential $V(\phi)$
driving the inflaton field. In the slow-roll limit, and for the 
relevant wavenumbers, the mass $m(t)$ is very approximately constant. 
We can therefore use the standard quantum mechanical treatment
for the harmonic oscillator \cite{qm} to find the wave function
for $\phi_{\bf k}$ (hereafter abbreviated as 
$\psi\equiv \psi(\phi_{\bf k})$).

In general $\psi$ may be written as a superposition of energy eigenmodes:
\be\label{psiexp}
\psi=\sum \alpha_n\psi_n
\ee
where $n$ labels the energy spectrum $E=\hbar \omega (n+1/2)$.
The $\psi_n$ take the form
\be
\psi_n(\phi)=C_nH_n{\left(\phi\over
{\sqrt 2} \sigma_0\right)}e^{-{\phi^2\over 4 \sigma_0^2}}
\ee
with normalisation fixing $C_n$ as,
\be
C_n={1\over (2^n n!{\sqrt{2\pi}}\sigma_0)^{1/2} }\, .
\ee
Here $\sigma_0^2$ is the variance  associated with the (Gaussian) probability
distribution for the ground state $|\psi_0|^2$. In other words
$\sigma_0^2$ is the power spectrum as derived in standard calculations
in which the field is assumed to be in the ground state. $\sigma_0$
is therefore 
related to the potential $V(\phi)$ and to $m$ according to the
standard formulae \cite{nv,pi}. We shall work with Hermite
polynomials $H_n(x)$ defined as 
\be\label{herm}
	H_n(x) = (-1)^n e^{x^2} \frac{d^x}{d x^n}e^{- x^2}
\ee
and normalised as
\be\label{ortho}
\int^\infty_{-\infty} e^{-x^2}H_n(x)H_m(x)dx=2^n \pi^{1/2} n!\delta_{nm}.
\ee
The most general probability density for the fluctuations 
in $\phi$ is thus $P=|\psi|^2$, with $\psi$ given by (\ref{psiexp}). 
The ground state (or ``zero-point'') fluctuations are Gaussian,
but any admixture with other states will be reflected
in a non-Gaussian distribution function \cite{nv,lesg}. 

It is worth explaining the apparent discrepancy between our expressions
and those in \cite{lesg}.
In \cite{lesg} the authors regard each mode $\phi_k$ as a complex variable,
and find a wave function for its modulus and phase, such that the probability distribution is most naturally expanded in 
Laguerre polynomials. In contrast, we consider the real and imaginary parts of $\phi_k$ as independent
real variables, and find separate wavefunctions for each. In this
way we find an expansion in Hermite polynomials, allowing for 
a simpler connection with the Edgeworth expansion.
The two treatments are equivalent and the probability
densities obtained in \cite{lesg} can be re-expressed in terms of
Hermite polynomials, and vice versa. The results in this Letter still
apply, although the algebra is significantly more tedious. 

\section{The Edgeworth expansion}
\label{edge}
We now review the Edgeworth expansion
\cite{Kendall}. This expansion is a particular form of a class of series
known as the Gram-Charlier Type A series.  These are used to construct
a distribution $P(x)$ by means of its $n$-th order moments $\mu_n$ 
or cumulants $\kappa_n$. These series
are convergent under  general conditions for $P(x)$ 
\cite{Kendall,Cramer}. 

For a standardized variable $x$ (ie $\langle x\rangle =0$ and
$\langle x^2\rangle =1$) the Edgeworth expansion takes the form
\be
P(x)={e^{-x^2/2}\over {\sqrt{2\pi}}}\sum_n c_n{\overline H}_n(x)\, .
\ee
If $x$ is not standartized this expression can be easily modified.
The coefficients $c_n$ are tabulated polynomials in the moments
$\mu_i$ or in the cumulants $\kappa_i$. Unfortunately the large body of
complex formulae available in the literature 
uses Hermite polynomials ${\overline H}_n$ defined as
\be
	{\overline H}_n(x) = (-1)^n e^{x^2/2} \frac{d^n}{dx^n}e^{-x^2/2}
\ee
as opposed to $H_n(x)$ defined in (\ref{herm}).
The two conventions
may be bridged by means of \cite{abra}
\be
{\overline H}_n(x)={H_n(x/{\sqrt 2})\over 2^{n/2}}
\ee

If $P(x)$ is expanded around a Gaussian with the correct variance 
$\sigma^2={\langle x^2\rangle}$ (with ${\langle x\rangle}=0$)
then the Edgeworth expansion is:
\ba \label{pofx}
P(x) &=& \frac{e^{{x^2\over 2\sigma^2}}}{\sqrt{2\pi}\sigma^2} 
{\Big (}
1+\frac{\kappa_3}{12{\sqrt 2}}H_3{\left(x\over {\sqrt 2}\sigma\right)}+
\frac{\kappa_4}{96}H_4{\left(x\over {\sqrt 2}\sigma\right)}
+\frac{\kappa_5}{480{\sqrt 2}}H_5{\left(x\over {\sqrt 2}\sigma\right)}
\nonumber\\
&&
+\frac{\kappa_6+10\kappa_3^2}{4960}H_6{\left(x\over {\sqrt 2}\sigma\right)}
+ ...{\Big )}
\ea
It is also possible to expand $P(x)$ around a Gaussian with a different
variance and mean.  The coefficients in the expansion are then
more complicated, but are listed in \cite{Kendall}.

\section{Coherent States}\label{res1}
Let us initially consider a {\it coherent} superposition of form 
(\ref{psiexp}). We first assume mild non-Gaussianity which we define
through the condition  $|\alpha_0|^2\gg |\alpha_i|^2$, for $i\ge1$. 
We can then consider the expansion of $|\psi|^2$ only 
to first order in $\alpha_i$,
\be
P(\phi)=|\psi|^2={e^{-{\phi^2\over 2 \sigma_0^2}}\over {\sqrt{2\pi}}
\sigma_0}{\left[1+\sum_{n\ge 1}{2\Re (\alpha_n)\over (2^n n!)^{1/2}}
H_n{\left(\phi\over {\sqrt 2}\sigma_0\right)}\right]}
\ee
where we have taken $\alpha_0$ to have zero phase (so that to
first order $\alpha_0=1$).
 
This leads to a one-to-one correspondence between the amplitudes
of the various energy eigenstates, and the combinations of cumulants
appearing as coefficients in the Edgeworth expansion. The latter simplify
enormously if we only keep first order terms: we find that
$\kappa_n\propto \Re(\alpha_n)$. Hence
to first order the coherent contamination of the ground state by 
the $n$\th energy eigenstate is signalled by a non-vanishing cumulant
$\kappa_n$. For instance
the presence of the third energy eigenstate results in 
$\kappa_3\propto \Re (\alpha_3)\ne 0$,
and, to first order, zero higher order cumulants. 

Of course we do not need to assume that $|\alpha_0|^2\gg |\alpha_i|^2$
for $i\ge 1$, but then the series' convergence will have to be 
more carefully watched.
For completeness, and for later use, we consider the more general case as well, making no assumptions about the relative amplitudes of the $\alpha_{i}$'s. We then have that,
\be 
P(\phi)=|\psi|^2={e^{-{\phi^2\over 2 \sigma_0^2}}\over {\sqrt{2\pi}}
\sigma_0}\sum_{i,j}{\alpha_i^*\alpha_j\over (2^{i+j} i!j!)^{1/2}}
H_i{\left(\phi\over {\sqrt 2}\sigma_0\right)}
H_j{\left(\phi\over {\sqrt 2}\sigma_0\right)}
\ee
We may recover the Edgeworth expansion by noting that, 
\be\label{expand}
    e^{-x^2} H_i(x)H_j(x)  = 
e^{-x^2} \left[ \sum_n b^{ij}_n H_n(x)\right]
\ee
with,
\be\label{bij}
	b^{ij}_n = 
\frac{2^{s-n}i!j!}{(s-n)!(s-i)!(s-j)!}
\ee
with $2s=n+i+j$. This expression is subject to two selection
rules: $n+i+j$ must be even, and the triangular inequality 
$|i-j|<n<i+j$. One may derive (\ref{bij})
using (\ref{ortho}) and the standard result for the
integral over a product of three Hermite polynomials (formula 7.375.2 
of \cite{Grad}). Thus we obtain the more complicated expression
\be 
P(\phi)=|\psi|^2={e^{-{\phi^2\over 2 \sigma_0^2}}\over {\sqrt{2\pi}}
\sigma_0}\sum_n{\left(\sum_{i,j}{b^{ij}_n
\alpha_i^*\alpha_j\over (2^{i+j} i!j!)^{1/2}}\right)}
H_n{\left(\phi\over {\sqrt 2}\sigma_0\right)}
\ee
Hence the coherent admixture of the $i$\th and $j$\th  eigenstates leads
to terms in the Edgeworth expansion to all orders between $|i-j|$
and $i+j$. Note that in general one has $\langle\phi\rangle\ne 0$,
and that the variance of $\phi$ may be different from $\sigma_0$,
requiring suitable modifications to the coefficients in the Edgeworth
expansion. 

\section{Incoherent States}\label{res2}
If the various energy eigenstates are added incoherently we obtain
the probability distribution,
\be
P(\phi)=|\psi|^2=\sum_n|\alpha_n|^2\psi_n^2=
{e^{-{\phi^2\over 2 \sigma_0^2}}\over {\sqrt{2\pi}}
\sigma_0}
\sum_{n}{|\alpha_n|^2\over 2^n n!}
H_n^2{\left(\phi\over {\sqrt 2}\sigma_0\right)}
\ee
As in the previous section this may be rewritten as
\be\label{edgeinc}
P(\phi)=
{e^{-{\phi^2\over 2 \sigma_0^2}}\over {\sqrt{2\pi}}
\sigma_0}\sum_k{\left(
\sum_{n}{b^{nn}_k|\alpha_n|^2\over 2^n n!}\right)}
H_k{\left(\phi\over {\sqrt 2}\sigma_0\right)}
\ee
and one may now use the Edgeworth expansion results to infer the
cumulants of the distribution. We see that in this case $\langle
\phi\rangle=0$. However the variance of the perturbed distribution
is never the same as $\sigma_0^2$ and is given by,
\be\label{varcor}
\sigma^2=\sigma_0^2\sum_n|\alpha_n|^2(2n+1)
\ee
a result which can be read off from (\ref{edgeinc}) (with the aid
of \cite{Kendall}), or easily
 derived in the Heisenberg picture. 
Hence formula (\ref{varcor}), providing $\sigma^2-\sigma_0^2$, must be used when solving for the cumulants of the distribution from the coefficients in (\ref{edgeinc}). 

The series (\ref{edgeinc}) only contains Hermite
polynomials of even order, since $b^{nn}_k$ can only be non-zero for
even $k$. Hence,
we note the important result that incoherent superpositions generate
non-Gaussian distributions with zero odd cumulants, in contrast with
what was obtained in the previous section (as
also mentioned in \cite{nv}).
Furthermore for incoherent superpositions
the presence of a non-vanishing $n$\th cumulant signals the
presence of the $n/2$\th energy eigenstate, in contrast with the
result for coherent superpositions. In general, even to lowest order
(which is second order in the amplitudes) one may not establish
a one-to-one relation between cumulants $\kappa_n$ and the
$|\alpha_n|^2$. The incoherent contamination of the ground state
by the $n$\th energy eigenstate generally leads to non-vanishing 
even cumulants of all orders up to $2n$.

\section{Discussion}
We have shown that standard inflation may generate density fluctuations
with the most general one-point distribution function,
if we allow the inflaton field not to be in the ground state. 
According to this scenario the cumulants of the distribution 
provide a measurement of the components of the wave function of the 
inflaton field, in terms of energy eigenstates. The explicit expression
depends on whether the superposition is coherent or incoherent. 
In the coherent case all cumulants can be potentially non-zero.
For incoherent superpositions only non-zero even cumulants can arise.

 
There is however one type of non-Gaussianity which cannot be generated
in this scenario: non-Gaussian inter-mode correlations.
Translational invariance requires that the two-point function for
modes $\phi_{\bf k}$ be diagonal ($\langle \phi^\star_{\bf k}
\phi_{{\bf k}'}\rangle =P_\phi(k)\delta ({\bf k}-{\bf k}')$).
If the $\phi_{\bf k}$ are Gaussian they are fully determined by
their two-point function, and so translational invariance
requires them to be independent random variables. 
However if the reduced higher order correlators {\it did} 
indeed exist they would need
not be diagonal as a 
result of translational  
invariance. For instance $\langle \phi^\star_{\bf k}
\phi_{\bf k}'\phi_{\bf k}''\rangle $ may be non-vanishing for all
modes satisfying ${\bf k}={\bf k}'+{\bf k}''$. Hence non-Gaussianity
allows inter-mode correlations, even if we impose translational 
invariance. These correlation cannot be generated
in the scenario we have proposed, because the modes $\phi_{\bf k}$
are genuine independent random variables.

One may wonder whether
the inflaton states we have considered are consistent with inflation. 
In most of the literature on inflationary density fluctuations
the inflaton is assumed to be in the ground state; two notable exceptions 
being \cite{nv} and \cite{lesg}.
This simplification has been justified using, amongst others, the rather dubious rationale that the field must be maximally symmetric, i.e. no scale should be privileged.
However scenarios have been considered where a preferred scale is present, in particular that the Planck scale itself is a
preferred scale \cite{nv}. 
A more physical argument for taking the ground state has also been proposed:
that there must be a cutoff in the energy spectrum of the resulting
inflaton particles to avoid infinite energy densities~\cite{lyth}. 
The cutoff must also be low enough so that the 
energy density of the inflaton particles does not dominate over the potential
driving inflation. The value of this cutoff is
uncomfortably low if the inflaton particles were ever in thermal
equilibrium. However it could happen that inflation is not
preceded by a thermal phase, but starts straight out of quantum
cosmology. In such a scenario one should not rule out mild
admixtures of states above the ground state, with ``mild'' being defined
in the sense that any backreaction must be negligible. 

It is of course debatable whether or not such superpositions
of states can be justified by quantum cosmology.
The Hartle and Hawking no-boundary proposal \cite{hartle,jonathan} 
and the Vilenkin tunneling wave proposal \cite{vilenkin} imply 
that the inflaton wave function must be in the ground 
state. It would be interesting to investigate 
the implications to quantum cosmology of a detection of non-Gaussianity.

\section*{Acknowledgments}
We thank Andy Albrecht, Pedro Ferreira, 
and Jonathan Halliwell for useful comments. 
We acknowledge financial support from the Beit Fellowship for
Scientific Research (CRC), PPARC (RB) and the Royal Society (JM).


\begin{thebibliography}{99}
\bibitem{guth81}A. H. Guth, {\sl Phys. Rev.} {\bf D23}, 347 (1981).
\bibitem{linde}A. Linde,  {\it Phys. Lett} {\bf B 108}, 1220 (1982).
\bibitem{as}A. Albrecht and P. Steinhardt, 
{\it Phys.Rev.Lett.} {\bf 48} 1220 (1982).
\bibitem{linde1}A. Linde, {\it Phys. Lett} {\bf B 129}, 177 (1983).
\bibitem{salopek}Salopek, D. S. {\sl Phys. Rev.} {\bf D45}, 
1139--1157  (1992).
\bibitem{mukh}A. Linde, V. Mukhanov
{\sl Phys. Rev.} {\bf D56}, (1997) 535-539.
\bibitem{peebles}P. J. E. Pebles, {\sl astro-ph/9805194;
astro-ph/9805212}.
\bibitem{pierpaoli99}E. Pierpaoli, J. Garcia-Bellido, S. Borgani; {\sl hep-ph/9909420}.
\bibitem{us}
P.G. Ferreira, J. Magueijo, and K.M.  G\'orski,
{\sl Astrophys. J. Lett.} {\bf 503}, L1--L4 (1998).
\bibitem{nov} D. Novikov, H. Feldman, S. Shandarin,
astro-ph/9809238.
\bibitem{nvs}M. Kamionkowski and A. Jaffe, {\it Nature} {\bf 395}
639 (1998).
\bibitem{pando}J. Pando, D. Valls-Gabaud, L.-Z. Fang,
  {\sl Phys. Rev. Lett.} {\bf 81} (1998) 4568-4571.
\bibitem{teg} B. Bromley and M. Tegmark, {\sl astro-ph/9904254}. 
\bibitem{band}A.J. Banday, 
S. Zaroubi, and K.M G\'orski,  {\sl astro-ph/9908070}. 
\bibitem{int}J. Magueijo, ``New non-Gaussian feature in COBE-DMR Four 
Year Maps'', Imperial/TP/98-99/071.
\bibitem{qm}E. Merzbacher, {\it Quantum Mechanics}, Wiley, NY, 1970. 
\bibitem{hartle}J. B. Hartle, S. W. Hawking, {\sl Phys. Rev.} {\bf D28},
2960 (1983).
\bibitem{jonathan}J.J. Halliwell adn S. W. Hawking, Phys. Rev. D {\bf 31}
1777 (1985).
\bibitem{nv}J. Martin, A. Riazuelo, M. Sakellariadou; {\sl astro-ph/9904167}.
\bibitem{lesg}J. Lesgourgues, D. Polarski, A. A. Starobinsky, {\it
Nucl.Phys.} {\bf B497} (1997) 479-510.
\bibitem{Kendall}Kendall, M.G. and Stuart, A.,
{\it The Advanced Theory of Statistics}, Charles Griffin (1977).
\bibitem{pi} A. Guth and S.-Y. Pi, {\sl Phys. Rev.} {\bf D32}, 
1899--1920  (1985).
\bibitem{barrow} J. Barrow and P. Coles, {\it M.N.R.A.S.} {\bf 244}
188 (1990).  
\bibitem{Cramer} J. M. Chambers, {\it Biometrika} (1967), {\bf 57},
367.
\bibitem{abra} M. Abramowitz and I. Stegun, Handbook of mathematical
formulae, Dover Publications, New York, 1972.
\bibitem{Grad} Gradshteyn, I. S., Ryzhik, I. M., {\it Table of
Integrals, Series, and Products}, Academic Press, (1996).
\bibitem{lyth}A. R. Liddle, D. H. Lyth, {\sl Phys. Rep} {\bf 231} 1 (1993).
\bibitem{vilenkin}A. Vilenkin, {\sl Phys. Lett.} {\bf B 117} 25 (1982).  
\end{thebibliography}
\end{document}